\documentclass{article}

\usepackage{arxiv}

\usepackage[utf8]{inputenc}
\usepackage[T1]{fontenc}
\usepackage{hyperref}
\usepackage{url}
\usepackage{booktabs}
\usepackage{amsfonts}
\usepackage{nicefrac}
\usepackage{microtype}
\usepackage{graphicx}
\usepackage{amsmath,amssymb}
\usepackage{bm}
\usepackage{mathrsfs}
\usepackage{amsthm}
\usepackage{multirow}
\usepackage{float}
\usepackage{xcolor}

\newtheorem{theorem}{Theorem}
\newtheorem{corollary}{Corollary}
\newtheorem{proposition}{Proposition}

\newcommand{\ZZ}{\mathbb{Z}_2}
\newcommand{\avg}[1]{\left\langle #1 \right\rangle}
\newcommand{\dd}{\mathrm{d}}
\newcommand{\ee}{\mathrm{e}}
\newcommand{\ii}{\mathrm{i}}
\newcommand{\ndot}{\dot{n}}
\newcommand{\Phisec}{\hat{\Pi}_{\Phi}}

\newcommand{\nbar}[1]{\bar{n}_{m_{#1}}}
\newcommand{\Psip}{|\Psi^{+}_{\ell_1\ell_2}\rangle}
\newcommand{\Phip}{|\Phi^{+}_{\ell_1\ell_2}\rangle}

\title{Predicting Noise in a Trapped-ion Quantum Simulation of a Lattice Gauge Theory}

\author{
  Hassan Ugail \\
  Centre for Visual Computing and Intelligent Systems \\
  University of Bradford \\
  United Kingdom \\
   \\
}

\begin{document}
\maketitle

\begin{abstract}
A recent trapped-ion experiment realised a lattice gauge theory in which the internal states of the ions carry the gauge field and their vibrations carry bosonic matter. In that experiment, a charge moving around a loop was frozen in place by Aharonov--Bohm interference. The noise of such a device is usually simulated after the measurements have been made. Here we show that it can be predicted before the machine is run. For each quantity that the experiment measures, we derive, from the reported noise rates alone, a formula for how it decays together with a bound it can never exceed. The results reveal a simple pattern. The local conservation law of the theory is a parity of the vibrational mode, so heating violates it at a rate that grows as the mode warms, and the violation accelerates rather than accumulating steadily. At zero electric field, each of the three noise processes of the device leaves its own fingerprint. Heating affects only the conservation law, motional dephasing only the interference, and qubit errors only the stored flux and the frozen state. Once the electric field is switched on, the frozen state begins to leak, and we write an exact formula and a guaranteed bound for that leak. 
\end{abstract}


\section{Introduction}
\label{sec:intro}

A gauge theory ties the motion of matter to the state of a field that lives on the links between lattice
sites. The local conservation law that expresses this tie is Gauss's law, and it is the quantity by which any
quantum simulation of a gauge theory must be judged~\cite{wiese,dalmonte,banuls}. The simplest discrete
example is the $\ZZ$ lattice gauge theory coupled to dynamical
matter~\cite{wegner,kogut1975,kogut1979,zohar,barbiero,borla}. Over the past decade, it has moved from a
theoretical model to a laboratory object. The first trapped-ion simulation of a gauge theory~\cite{martinez} followed earlier
proposals~\cite{hauke2013}. The $\ZZ$ theory itself has since been realised with ultracold atoms in
Floquet-engineered double wells~\cite{schweizer,gorg} and with superconducting qubits in which both matter and
gauge field are spins~\cite{mildenberger}. Most recently, it has been realised on a trapped-ion processor in
which the gauge field is encoded in the internal state of each ion and the matter in the quantised vibrations
of the crystal~\cite{bazavan,saner}.

The hybrid qubit--oscillator encoding has two features that matter here. A matter site can hold any number of
excitations, so the matter is bosonic rather than hard-core. Synthetic dimensions built from qubit--oscillator
couplings allow loop geometries to be assembled from a small number of ions. The experiment of Saner and
co-workers used two ions to close a two-link loop, prepared the link qubits in Bell states that encode a $\ZZ$
flux through the loop, and found that a phonon tunnels freely when the flux is zero and is frozen in place by
destructive interference when the flux is $\pi$~\cite{saner}. This is the Aharonov--Bohm
effect~\cite{aharonov} with a dynamical gauge field, and the frozen state is the gauge-theory counterpart of an
Aharonov--Bohm cage~\cite{vidal,domanti}.

Every such device is noisy, and the experiment reported its noise with care. The vibrational modes heat at a
few to a few hundred quanta per second, their motional coherence times are a few milliseconds, and the gauge
qubits are prepared with a Bell-state fidelity below one. The usual practice is to include these rates in a
Lindblad simulation of the full dynamics and to compare the result with the data. That practice tells us
whether the device behaved as modelled. It does not tell an experimenter who is planning the next geometry how
much each observable can be trusted before the machine is run, nor which noise process is responsible when an
observable cannot be trusted. The theory of gauge violation under noise has addressed part of this question.
Halimeh and Hauke showed that coherent gauge-breaking errors produce a violation that grows quadratically in
time from a gauge-invariant state~\cite{halimeh2020reliability}. With Kasper they showed that Markovian
decoherence produces a violation that grows linearly, so that the two mechanisms can be told apart by their
short-time behaviour~\cite{halimeh2020fate,halimeh2020crossover}. A programme of gauge protection by energy
penalties has grown from this work~\cite{halimeh2021protection,stannigel}. These studies treat hard-core
bosons or spins as matter. In the ion platform the matter is a soft boson, and the Gauss generator at a site
is the phonon parity of that mode multiplied by the qubits on the links that meet it. Heating flips this
parity, and it flips it at a rate that depends on how many quanta the mode already holds. That single fact is
the starting point of the present work.

We ask, for each of the three observable classes of the experiment, whether its degradation under each noise
channel can be written in closed form from the channel description, and whether the closed form can be
accompanied by a bound that is guaranteed rather than estimated. The three classes are the Gauss generators,
the loop fluxes, and the populations in which interference is read off. We organise the answers in the
three-tier form that we earlier introduced for the certification of Weyl diagnostics in numerical
relativity~\cite{ugail2026weyl}. A certified bound is a statement that cannot be violated. A first-order law is
a formula that is usually accurate. A calibrated gate is a rule, tested on random instances with a declared
coverage, that says when the law may be trusted. The same architecture underlies our work on predicting the
behaviour of noisy quantum neural networks from the description of their noise
channels~\cite{ugail2026access,ugail2026coherence}, and this paper carries that programme from quantum machine
learning to lattice gauge theory.

Our results fall into three groups. For Gauss's law, we prove an exact equation of motion under heating whose
rate carries the factor $2\hat n_m+1$. We turn it into a certified bound by using the fact that the total
phonon number is conserved apart from a known linear ramp, and we extract a first-order law from the noiseless
populations. The bound shows that the violation is superlinear in time, which hard-core matter cannot show,
and both the law and the bound hold on every geometry and at every electric field we tested. For the flux and
the frozen sector, we show that at zero electric field, the flux correlator and the protection of a $\pi$-flux
loop are untouched by motional heating, motional dephasing and qubit dephasing, and are broken only by qubit
bit flips, for which we write exact exponential laws. Once the electric field is switched on, the frozen sector
leaks coherently. We show that the leak has an exact closed form on a three-level system, and that a
certificate for it can be assembled on small phonon-number sectors without a full simulation. For the
interference contrast, we show that at zero field heating is invisible to the contrast for any matter state,
including squeezed matter, because the dynamics is linear in the bosons, while motional dephasing sets the
contrast decay at the damped two-site rate. With the field on, the heating shift of the populations becomes
non-additive at the level of the heating effect itself. We quantify and bound this effect, and we leave its
closed form open.

The paper therefore offers three things to the gauge-theory community. It gives the exact form, for soft
bosonic matter, of the incoherent gauge-violation law whose leading order is due to Halimeh, Kasper and
Hauke~\cite{halimeh2020fate,halimeh2020crossover}. It gives the noise robustness of the $\pi$-flux walls whose
coherent physics is developed by Domanti, Bermudez and Amico~\cite{domanti}. And it shows that the three noise
channels of the ion device have disjoint fingerprints at zero field, together with a precise account of where
that separation fails. All three are stated as certified bounds with predictive laws beside them, so that an
experimenter can use them before a run rather than after it.

\section{Model, noise channels and observables}
\label{sec:model}

We consider a graph whose sites $m$ carry oscillators with ladder operators $\hat a_m$ and whose links
$\ell=(i,j)$ carry qubits with Pauli operators $\hat\sigma^{x}_\ell$, $\hat\sigma^{y}_\ell$ and
$\hat\sigma^{z}_\ell$. The Hamiltonian of the $\ZZ$ lattice gauge theory on this graph is
\begin{equation}
\hat H=\sum_{\ell=(i,j)}\Big[J\big(\hat a_j^{\dagger}\hat\sigma^{z}_\ell\hat a_i+\mathrm{h.c.}\big)
+h\,\hat\sigma^{x}_\ell\Big],
\label{eq:H}
\end{equation}
where $J$ is the gauge-invariant tunnelling and $h$ is the electric field~\cite{bazavan,saner}. A phonon hops
from site $i$ to site $j$ only through the link qubit, which it flips in the electric basis, and the electric
term costs energy for every link that carries an electric line. The generator of the local symmetry at site
$m$ is
\begin{equation}
\hat G_m=\hat P_m\prod_{\ell\ni m}\hat\sigma^{x}_\ell,\qquad \hat P_m=\ee^{\ii\pi\hat a_m^{\dagger}\hat a_m},
\label{eq:G}
\end{equation}
and $[\hat H,\hat G_m]=0$ for every site. The eigenvalue $g_m=\pm1$ of $\hat G_m$ labels the background
charge at $m$. Gauss's law is the statement that the dynamics stays in the sector fixed by the initial state.
For a closed loop $C$ of links the flux is measured by the Wilson operator $\hat
W_C=\prod_{\ell\in C}\hat\sigma^{z}_\ell$. The experiment of~\cite{saner} is the case of two sites joined by two
links. Preparing the link qubits in $\Phip$ or $|\Phi^{-}_{\ell_1\ell_2}\rangle$ fixes the flux to zero, and
preparing them in $\Psip$ or $|\Psi^{-}_{\ell_1\ell_2}\rangle$ fixes it to $\pi$. We write
$\Phisec=\tfrac12(1+\hat\sigma^{z}_{\ell_1}\hat\sigma^{z}_{\ell_2})$ for the projector onto the zero-flux
sector of such a loop.

The device noise is modelled by a Lindblad master equation $\dot\rho=-\ii[\hat H,\rho]+\sum_k\mathcal D[\hat
L_k]\rho$ with $\mathcal D[\hat L]\rho=\hat L\rho\hat L^{\dagger}-\tfrac12\{\hat L^{\dagger}\hat L,\rho\}$
\cite{lindblad,gks,breuer}. Motional heating of mode $m$ at a rate of $\ndot_m$ quanta per second is modelled
as an infinite-temperature bath with jump operators $\sqrt{\ndot_m}\,\hat a_m$ and $\sqrt{\ndot_m}\,\hat
a_m^{\dagger}$, which gives $\dd\avg{\hat n_m}/\dd t=\ndot_m$ exactly in the absence of other
dynamics~\cite{turchette,brownnutt}. Motional dephasing with coherence time $t_{c,m}$ is modelled by the jump
operator $\sqrt{2/t_{c,m}}\,\hat a_m^{\dagger}\hat a_m$, so that a superposition of the zero- and one-phonon
states loses its coherence at the rate $1/t_{c,m}$. Qubit errors on link $\ell$ are modelled as bit flips
$\sqrt{\gamma_x}\,\hat\sigma^{x}_\ell$ and as dephasing $\sqrt{\gamma_z}\,\hat\sigma^{z}_\ell$.

The experiment quotes heating rates of $300$, $5$ and $17$ quanta per second and coherence times of $1.7$,
$2.3$ and $3.3$ milliseconds for its three modes, an initial thermal occupation $\bar n_0=0.1$ on every mode,
a Bell-state fidelity $\mathcal F=0.943$, and tunnelling rates $J/\pi$ of $1.47$ and $1.51$ kilohertz on the
link and $0.35$ kilohertz on the loop~\cite{saner}. It does not quote qubit error rates, so $\gamma_x$ and
$\gamma_z$ are carried as free parameters. We use the radial-mode values $\ndot=5$ and $17$ quanta per second
and $t_c=2.3$ and $3.3$ milliseconds throughout the loop and chain calculations, and we model the Bell-state
infidelity as the mixture $\mathcal F\,|\mathrm{Bell}\rangle\langle\mathrm{Bell}|+(1-\mathcal F)\,\mathbb
I/4$.

Three classes of observable are studied. The first is the Gauss generator $\avg{\hat G_m}$ at each site,
which registers the violation of the local conservation law. We also report the violation in the convention of
Halimeh and Hauke, $\varepsilon_m=1-g_m\avg{\hat G_m}$, so that the results of Section~\ref{sec:gauss} can be
compared with the hard-core-boson literature. The second is the flux correlator $\avg{\hat W_C}$, written
$\bar s^{zz}_{\ell_1\ell_2}$ on the two-link loop in the notation of~\cite{saner}, together with the zero-flux
sector population $\avg{\Phisec}$ when the system is prepared in a $\pi$-flux state. The third is the set of
matter populations $\nbar{i}=\avg{\hat a_i^{\dagger}\hat a_i}$ and gauge correlators $\bar
s^{xx}_{\ell_1\ell_2}$ in which interference is read off. All calculations are exact Lindblad evolutions in a
truncated Fock space using QuTiP~\cite{qutip}. The truncations are chosen so that the results are independent
of them to the precision quoted, and every run is deterministic under a fixed seed. The details are given in
Appendix~\ref{app:numerics}.

\section{Gauss's law under heating}
\label{sec:gauss}

The Gauss generator at a site is the parity of the matter mode dressed with the link qubits. Among the noise
channels of the device, motional dephasing is a function of the phonon number and therefore commutes with the
parity. Qubit dephasing acts with $\hat\sigma^{z}$ and therefore anticommutes with each $\hat\sigma^{x}$
factor. Heating adds or removes one phonon and therefore flips the parity. The Hamiltonian commutes with $\hat
G_m$ by construction. These four facts fix the equation of motion of the Gauss generator exactly.

\begin{theorem}[Gauss's law under heating]
\label{thm:gauss}
For any graph, any electric field $h$ and any matter state, under motional heating at rates $\ndot_m$,
motional dephasing at any rates and qubit dephasing at rate $\gamma_z$ on each link,
\begin{equation}
\frac{\dd}{\dd t}\avg{\hat G_m}=-2\ndot_m\,\avg{\hat G_m\,(2\hat n_m+1)}-2\gamma_z d_m\avg{\hat G_m},
\label{eq:gaussode}
\end{equation}
where $d_m$ is the number of links that meet site $m$.
\end{theorem}

The proof is a short computation of the adjoint Lindbladian. Using $\hat a\hat P=-\hat P\hat a$ and $\hat
a^{\dagger}\hat P=-\hat P\hat a^{\dagger}$, the jump $\sqrt{\ndot}\,\hat a$ gives $\mathcal L^{\dagger}[\hat
P]=\ndot(\hat a^{\dagger}\hat P\hat a-\hat P\hat a^{\dagger}\hat a)=-2\ndot\hat P\hat n$, and the jump
$\sqrt{\ndot}\,\hat a^{\dagger}$ gives $-2\ndot\hat P(\hat n+1)$. Together they give $\mathcal L^{\dagger}[\hat
P]=-2\ndot\hat P(2\hat n+1)$. The dephasing jump $\hat n$ commutes with $\hat P$ and contributes nothing, the
Hamiltonian contributes nothing because it commutes with $\hat G_m$, and each qubit dephasing jump on a link
at $m$ contributes $-2\gamma_z\hat G_m$. We set $\gamma_z=0$ in what follows, since the qubit dephasing term is
a simple exponential factor.

The theorem has two useful consequences. Since $\hat G_m$ commutes with $\hat n_m$ and $\hat G_m^2=1$, the
operator $\hat G_m(2\hat n_m+1)$ is Hermitian with modulus $2\hat n_m+1$, so that $|\avg{\hat G_m(2\hat
n_m+1)}|\le\avg{2\hat n_m+1}$. The occupation of a single mode is bounded by the total phonon number $\hat
N=\sum_m\hat n_m$. The hopping and the dephasing conserve $\hat N$ and the heating adds $\sum_m\ndot_m$ quanta
per second, so the expectation $\avg{\hat N}(t)=N_0+\sum_m\ndot_m t$ is known exactly at all times.

\begin{corollary}[Certified bound]
\label{cor:gausscert}
With $N_0$ the initial total phonon number,
\begin{equation}
\big|\avg{\hat G_m}(t)-\avg{\hat G_m}(0)\big|\le
2\ndot_m\Big[(2N_0+1)\,t+\tfrac12\Big(\sum_{m'}\ndot_{m'}\Big)t^2\Big].
\label{eq:gausscert}
\end{equation}
\end{corollary}

The $t^2$ term is the signature of soft matter. As the mode heats, the rate at which its parity flips
increases, and the violation of Gauss's law accelerates. For hard-core bosons a loss event on an occupied site
flips the parity with probability one and nothing accumulates, so the violation grows linearly in time. That
is the diffusive law of~\cite{halimeh2020fate,halimeh2020crossover}. Equation~\eqref{eq:gaussode} reduces to it
at leading order and short times, and departs from it as soon as the occupation changes.

\begin{corollary}[First-order law]
\label{cor:gausslaw}
Replacing the correlator in Eq.~\eqref{eq:gaussode} by the product of its expectations, and the noisy
population by the noiseless population plus the heating ramp,
\begin{equation}
\avg{\hat G_m}(t)\approx\avg{\hat G_m}(0)\exp\Big[-2\ndot_m\int_0^t\big(2\nbar{}^{(0)}(s)+2\ndot_m s+1\big)\dd s\Big],
\label{eq:gausslaw}
\end{equation}
where $\nbar{}^{(0)}(s)$ is the population of mode $m$ under the noiseless dynamics.
\end{corollary}

The first-order law needs no simulation in practice, because the noiseless population on the link and on the
loop is known in closed form~\cite{bazavan}. It omits the correlation between the parity and the occupation.
That correlation is negative, because states whose parity has flipped carry more quanta on average, and the
law therefore overestimates the decay once $\avg{\hat G_m}$ is small.

\begin{figure}
\centering
\includegraphics[width=0.98\textwidth]{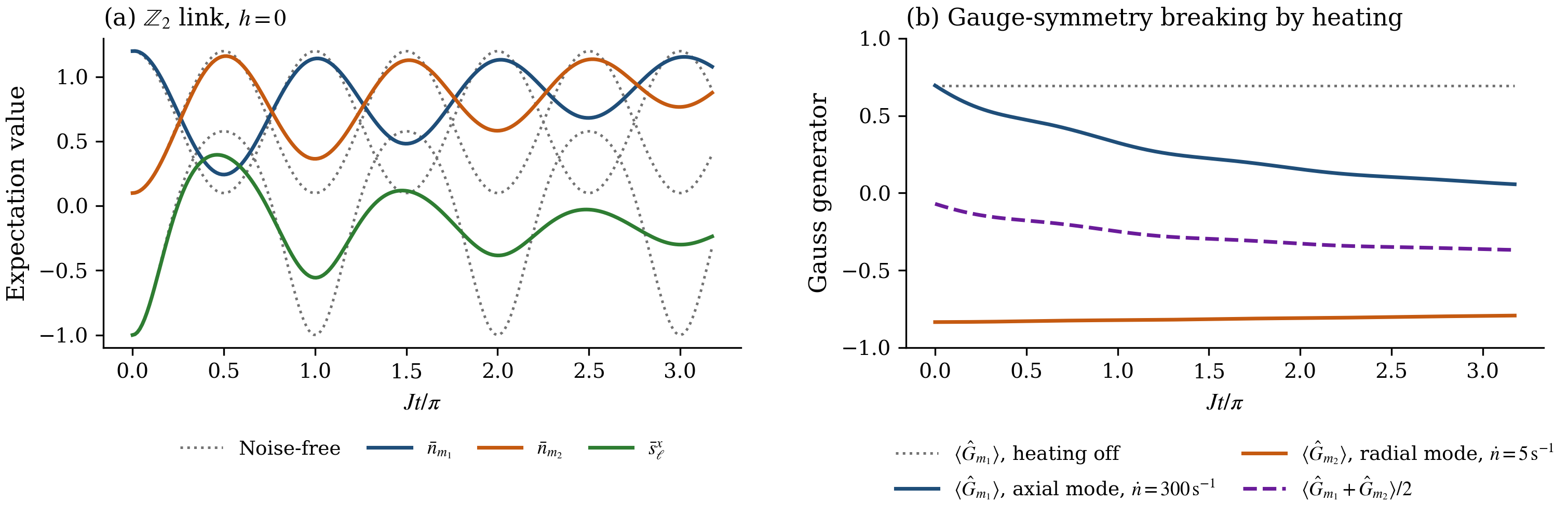}
\caption{The single $\ZZ$ link of the experiment at $h=0$ with the stated noise parameters. (a) Matter
populations $\nbar{1}$ and $\nbar{2}$ and the gauge field $\bar s^{x}_{\ell}$ under full motional noise,
with the noise-free curves dotted. (b) The two Gauss generators. The generator at the axial mode, which heats
at $300$ quanta per second, falls from $0.69$ to $0.06$ within the measurement window, while the generator at
the radial mode, which heats at $5$ quanta per second, barely moves. With the heating switched off the
generator is exactly constant.}
\label{fig:link}
\end{figure}

Figure~\ref{fig:link} shows the single link of the experiment with its stated noise. This is the only
calculation in the paper that uses the axial mode, whose heating rate of $300$ quanta per second is sixty
times that of the cold radial mode, and it is included because it shows how unequal the gauge-symmetry
breaking can be between two modes of the same device. The matter and gauge observables reproduce the familiar
correlated Rabi flopping~\cite{bazavan}. With the heating switched off the generator is constant to machine
precision, which is the statement that motional dephasing commutes with Gauss's law.

\begin{figure}
\centering
\includegraphics[width=0.56\textwidth]{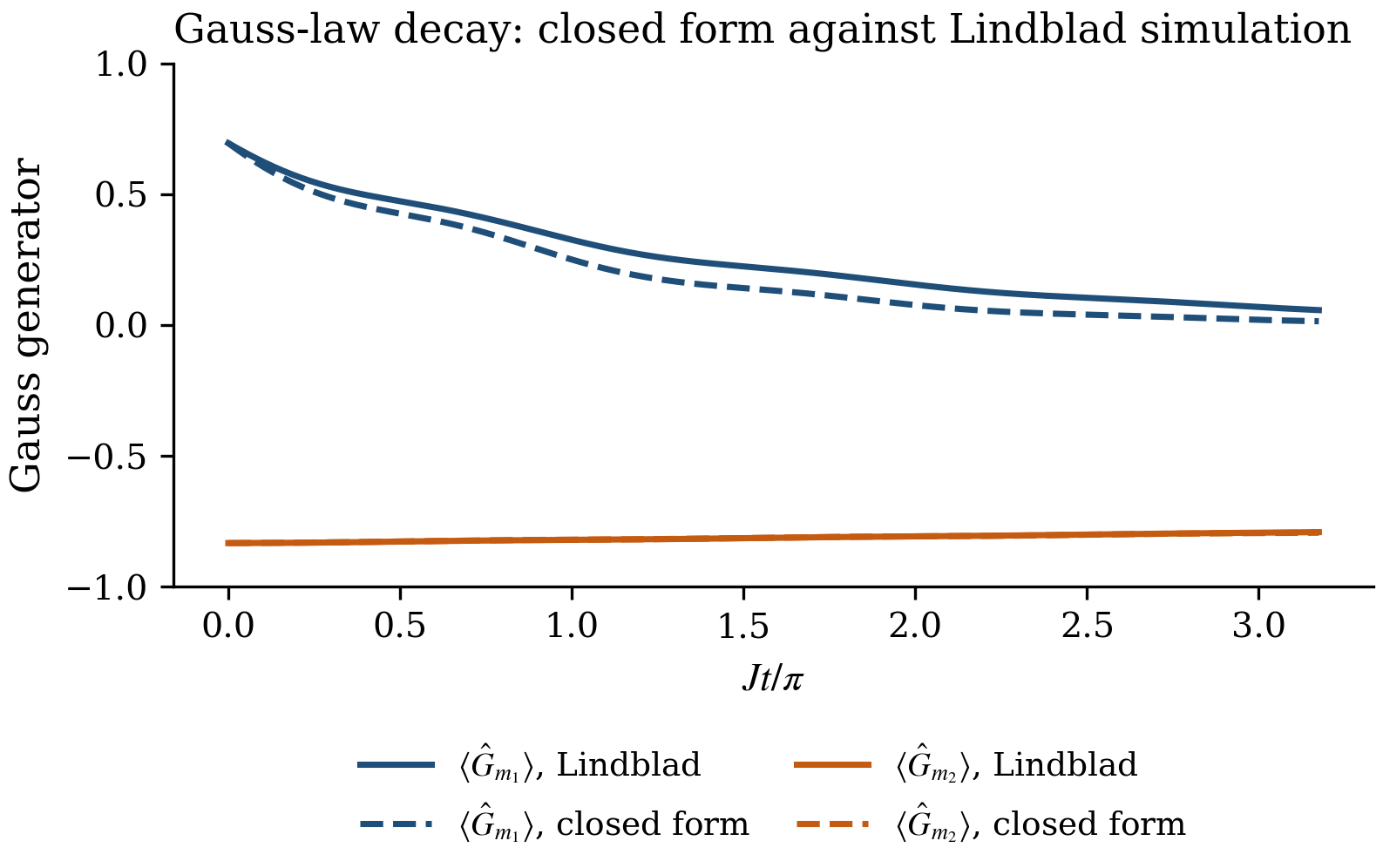}
\caption{Gauss-law decay on the link. Solid curves are the Lindblad evolution and dashed curves are the
first-order law of Eq.~\eqref{eq:gausslaw}. On the radial mode the law is accurate to $0.002$. On the axial
mode the omitted correlation between parity and occupation makes the law over-predict the decay once the
generator is small.}
\label{fig:gausslaw}
\end{figure}

Figure~\ref{fig:gausslaw} compares the first-order law with the Lindblad evolution on the same link. On the
cold radial mode the law is accurate to $0.002$ in absolute terms over the full window. On the hot axial mode
it is accurate to $0.08$, and the discrepancy has the sign that the omitted correlation predicts. Integrating
the exact equation~\eqref{eq:gaussode} with the simulated correlator reproduces $\avg{\hat G_m}$ to
$8\times10^{-4}$, which confirms that the equation is exact and that the error of the law lies entirely in the
factorisation.

The theorem makes no reference to the geometry or to the electric field, and we have tested it accordingly. On
the two-link loop at $h/J\in\{0,0.3,0.6,1.0\}$ the first-order law for the radial-mode generator has a maximum
absolute error between $0.016$ and $0.020$, with no trend in $h$, and the certified bound holds at every time
point. On a chain of two loops with three sites and four links, which is the first of the roadmap geometries
of~\cite{saner}, the middle site has degree four and its generator carries the product of four link qubits.
There the law is accurate to $0.008$ at $h=0$ and to $0.011$ at $h/J=0.3$, and the certified band is about
three times wider than the actual drift, as Figure~\ref{fig:chain}(c) shows. On a triangular loop the law is
accurate to better than $10^{-4}$ on the site that carries the cold mode. These tests span degree two and
degree four, zero and finite field, and a single phonon and a thermal mixture, and all of them agree with the
theorem.

\section{Flux, vison walls and Aharonov--Bohm protection}
\label{sec:flux}

The second class of observable concerns the gauge field itself. We first treat the flux and the frozen sector
at zero electric field, where every statement is exact, and then the frozen sector at finite field, where a
closed form and a certificate take the place of exactness.

\subsection{The flux and the frozen sector at zero field}
\label{sec:zerofield}

The flux operator $\hat W_C$ is a product of $\hat\sigma^{z}$ over the links of a loop. It commutes with the
hopping term of Eq.~\eqref{eq:H}, with every motional jump operator, since these act on the oscillators alone,
and with qubit dephasing. Under the noise model that the experiment states, the flux is therefore an exact
constant of motion at $h=0$. The only channel that can move it is a qubit error that does not commute with
$\hat\sigma^{z}$. For independent bit flips at rate $\gamma_x$ on each link, $\mathcal
L^{\dagger}[\hat\sigma^{z}_\ell]=\gamma_x(\hat\sigma^{x}\hat\sigma^{z}\hat\sigma^{x}-\hat\sigma^{z})=-2\gamma_x\hat\sigma^{z}_\ell$,
and the product over the $|C|$ links of the loop gives the following result.

\begin{theorem}[Flux law]
\label{thm:flux}
At $h=0$, under motional heating, motional dephasing, qubit dephasing and qubit bit flips at rate $\gamma_x$
per link,
\begin{equation}
\avg{\hat W_C}(t)=\avg{\hat W_C}(0)\,\ee^{-2\gamma_x|C|\,t}
\label{eq:fluxlaw}
\end{equation}
exactly. At $h\neq0$ the same expression is the first-order law for the noise-induced factor, and the
noise-induced deviation is certified by $|\Delta\avg{\hat W_C}|\le2\gamma_x|C|\,t$.
\end{theorem}

\begin{figure}
\centering
\includegraphics[width=0.98\textwidth]{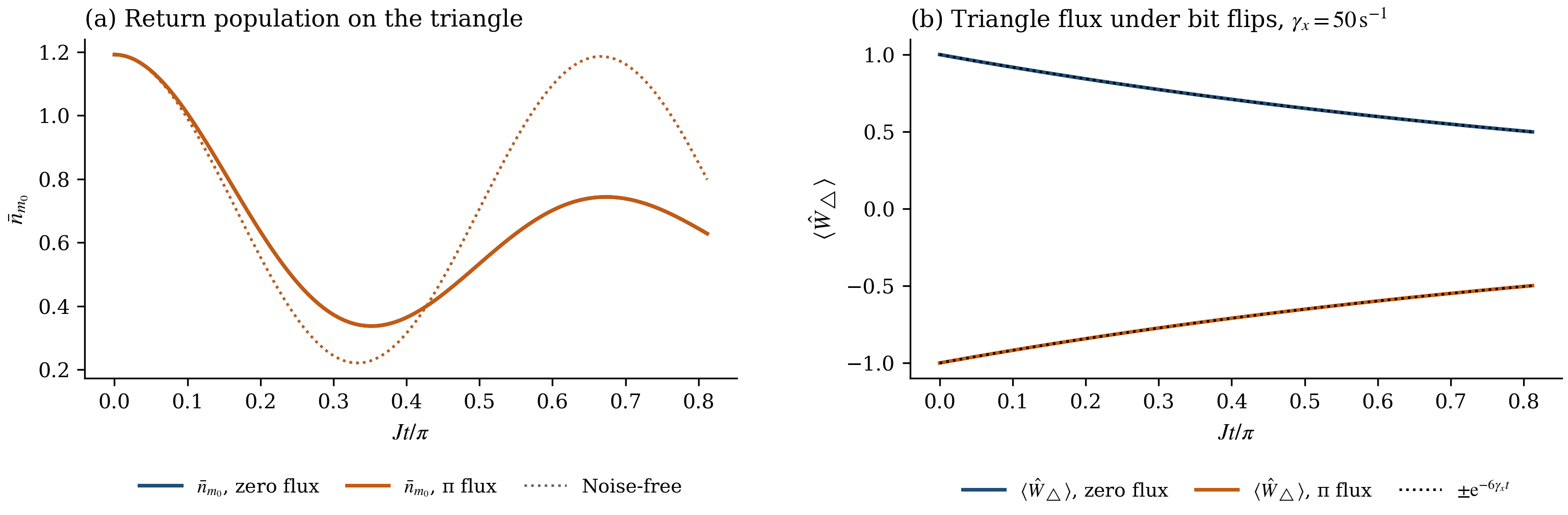}
\caption{The triangular loop with three sites and three links at $h=0$. (a) Return population on the initial
site for zero flux and $\pi$ flux under motional noise, with the noise-free curves dotted. A $\pi$-flux
triangle has no compact localised state, so there is no cage and only a changed return probability. (b) The
triangle flux under bit flips at $\gamma_x=50$ per second follows $\pm\ee^{-6\gamma_x t}$ to $2\times10^{-9}$.}
\label{fig:triangle}
\end{figure}

On the two-link loop the decay rate is $4\gamma_x$, and the simulation agrees with Eq.~\eqref{eq:fluxlaw} to
one part in $10^{9}$ for $\gamma_x$ between $25$ and $400$ per second. On the triangle the rate is $6\gamma_x$
and the agreement is $2\times10^{-9}$, as Figure~\ref{fig:triangle} shows. The result bears directly on the
experiment. The supplementary material of~\cite{saner} reports a slow decay of $\bar s^{zz}_{\ell_1\ell_2}$ and
attributes it to decoherence. Under the noise channels the paper lists, that correlator cannot decay at all.
The observed decay therefore measures a qubit error of bit-flip or depolarising type that the stated noise
model does not contain, and its slope gives $\gamma_x$ through Eq.~\eqref{eq:fluxlaw}.

We now turn to the frozen sector. On the two-link loop the hopping term is $J(\hat a_2^{\dagger}\hat
a_1+\mathrm{h.c.})(\hat\sigma^{z}_{\ell_1}+\hat\sigma^{z}_{\ell_2})$, and the operator
$\hat\sigma^{z}_{\ell_1}+\hat\sigma^{z}_{\ell_2}$ annihilates both $\Psip$ and $|\Psi^{-}_{\ell_1\ell_2}\rangle$.
The hopping therefore annihilates every state whose gauge part lies in the $\Psi$ sector, whatever the matter
does. This is the algebraic content of the freezing observed in~\cite{saner}, and it is the mechanism by which
a $\pi$-flux loop acts as a wall in a chain of loops~\cite{domanti}. The question for a noisy device is which
channels can open the wall.

\begin{theorem}[Wall certificate at zero field]
\label{thm:wall}
Let the two-link loop be prepared with $\pi$ flux and $h=0$. Under motional heating, motional dephasing and
qubit dephasing at any rates, the zero-flux sector population $\avg{\Phisec}$ remains zero and no matter
crosses the loop. Under qubit bit flips at rate $\gamma_x$ per link,
\begin{equation}
\avg{\Phisec}(t)=\tfrac12\big(1-\ee^{-4\gamma_x t}\big)
\label{eq:leak0}
\end{equation}
exactly, and the population that crosses the loop is bounded by this quantity.
\end{theorem}

\begin{figure}
\centering
\includegraphics[width=0.98\textwidth]{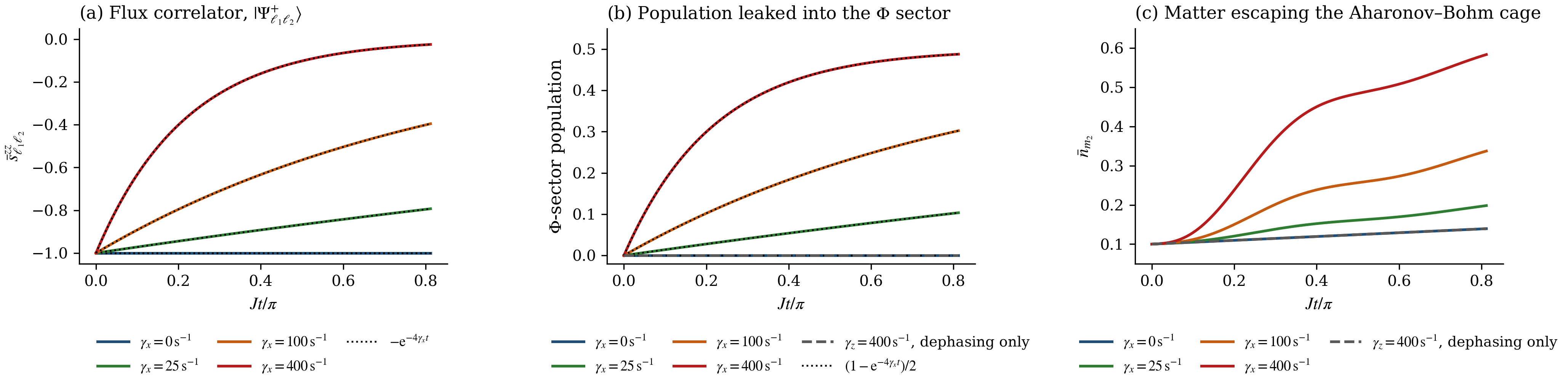}
\caption{The wall at zero field under qubit bit flips, with the two-link loop prepared in $\Psip$. (a) The
flux correlator follows $-\ee^{-4\gamma_x t}$ exactly. (b) The population leaked into the zero-flux sector
follows $(1-\ee^{-4\gamma_x t})/2$ exactly, while qubit dephasing at $400$ per second leaks nothing. (c) The
matter that escapes the cage. Motional noise and qubit dephasing alone leave the phonon frozen.}
\label{fig:leak}
\end{figure}

Heating changes the matter charge but leaves the gauge sector untouched, so the hopping continues to
annihilate the state. Qubit dephasing maps $\Psi^{+}$ to $\Psi^{-}$ and back, and both are frozen. Only a
single-qubit $\hat\sigma^{x}$ or $\hat\sigma^{y}$ error maps a $\Psi$ state to a $\Phi$ state, and once the
gauge field is in the $\Phi$ sector the matter tunnels at rate $2J$. The population in the $\Phi$ sector under
independent bit flips is the probability of an odd number of flips, which is Eq.~\eqref{eq:leak0}.
Figure~\ref{fig:leak} verifies the three statements. The flux correlator under bit flips follows
$-\ee^{-4\gamma_x t}$, the leaked population follows Eq.~\eqref{eq:leak0} to one part in $10^{9}$, and qubit
dephasing at $400$ per second leaks nothing.

\begin{figure}
\centering
\includegraphics[width=0.98\textwidth]{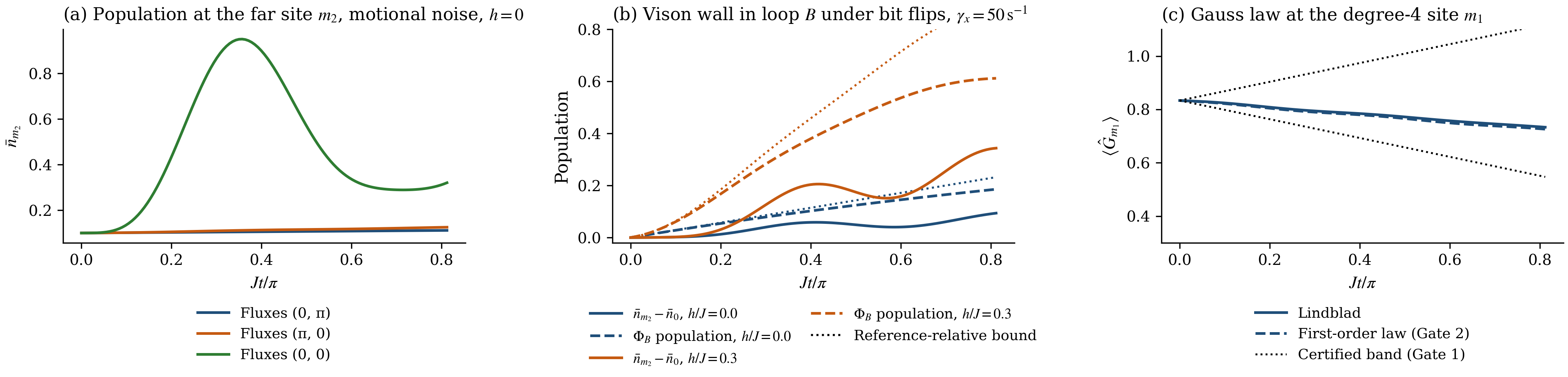}
\caption{The chain of two loops with three sites and four links under the radial-mode noise. (a) Population at
the far site for the three flux configurations at $h=0$. The phonon is blocked by a $\pi$-flux loop wherever
it stands. (b) The vison wall in the second loop under bit flips at $\gamma_x=50$ per second, at $h=0$ and at
$h/J=0.3$. The excess population at the far site stays below the $\Phi$-sector population of that loop, which
stays below the reference-relative bound of Section~\ref{sec:wallfield}. (c) The Gauss generator at the
degree-four middle site with the first-order law and the certified band.}
\label{fig:chain}
\end{figure}

On the chain of two loops the same argument applies loop by loop. With the fluxes $(0,\pi)$ a phonon that
starts at the first site reaches the middle site and is stopped at the second loop. Under bit flips at $50$ per
second the population at the far site never exceeds the $\Phi$-sector population of the second loop, which in
turn never exceeds Eq.~\eqref{eq:leak0}, as Figure~\ref{fig:chain}(b) shows.

\subsection{The frozen sector at finite field}
\label{sec:wallfield}

With the electric field on, the term $h(\hat\sigma^{x}_{\ell_1}+\hat\sigma^{x}_{\ell_2})$ maps $\Psip$ to
$\Phip$ with amplitude $2h$, and the wall opens coherently. The dynamics from $|1,\Psi^{+},0\rangle$ is
confined to three states. The hopping maps $|1,\Phi^{+},0\rangle$ to $|0,\Phi^{-},1\rangle$ with amplitude
$2J$, and the electric term annihilates $|\Phi^{-}_{\ell_1\ell_2}\rangle$, which closes the chain. This is
the $\Lambda$ system that B\u{a}z\u{a}van and co-workers identified in the two-boson sector of a single
link~\cite{bazavan}, here appearing in the one-boson sector of the loop. In the basis
$\{|1,\Psi^{+},0\rangle,|1,\Phi^{+},0\rangle,|0,\Phi^{-},1\rangle\}$ the Hamiltonian is the tridiagonal matrix
with off-diagonal entries $2h$ and $2J$ and zero diagonal, and its eigenvalues are $0$ and $\pm\Omega$ with
$\Omega=2\sqrt{h^2+J^2}$.

\begin{proposition}[Coherent leak]
\label{prop:lambda}
For a pure single phonon on the first site and the gauge field in $\Psip$, the zero-flux sector population
under Eq.~\eqref{eq:H} is
\begin{equation}
P^{\Lambda}_{\Phi}(t)=\frac{h^2}{h^2+J^2}\sin^2\Omega t+\frac{h^2J^2}{(h^2+J^2)^2}\big(1-\cos\Omega t\big)^2 .
\label{eq:lambda}
\end{equation}
\end{proposition}

The expression follows by expanding the initial state in the three eigenvectors and reading off the two
$\Phi$-sector amplitudes, as Appendix~\ref{app:sectors} records. It agrees with a direct exponentiation of the
three-by-three matrix to $10^{-15}$, and with the full Lindblad evolution of a pure single phonon in a Fock
space of dimension four per mode to better than $6\times10^{-6}$ at $h/J=0.3$, $0.6$ and $1.0$. At $h=J$ the
second term reaches one at $\Omega t=\pi$. The frozen sector then empties completely and periodically when the
electric field equals the tunnelling, which is a sharp prediction for the experiment.

To turn the closed form into a certificate we use two further structural facts. Motional dephasing is diagonal
in the phonon numbers and therefore keeps the dynamics inside each sector of fixed total phonon number. The
thermal initial matter state of the experiment is diagonal in the Fock basis. The dephasing-only dynamics
therefore decomposes exactly into a mixture of small sectors, each of dimension at most four times the number
of occupation pairs $(n_1,n_2)$ with $n_1+n_2=N$, and the decomposition does not depend on any Fock
truncation. We retain the sectors with $N\le3$, which carry $99.6$ per cent of the initial weight, and we bound
the remaining weight $w_{\rm rest}=0.0037$ by one. The resulting reference $P^{\rm ref}_{\Phi}(t)$ agrees with
the full dephasing-only simulation to between $5\times10^{-4}$ and $2.5\times10^{-3}$, which is the size of the
discarded weight.

Heating is the only channel that leaves these sectors, and bit flips act on the qubits alone. For a
Lindbladian split as $\mathcal L=\mathcal L_0+\mathcal D$, the Duhamel identity $\rho_t-\rho^{(0)}_t=
\int_0^t\ee^{(t-s)\mathcal L}\mathcal D\,\rho^{(0)}_s\,\dd s$ and the contractivity of completely positive
trace-preserving maps in trace norm~\cite{nielsen} give $\|\rho_t-\rho^{(0)}_t\|_1\le\int_0^t\|\mathcal
D\rho^{(0)}_s\|_1\dd s$. For a projector the change in expectation is at most half the trace distance. The
bit-flip dissipator on two links satisfies $\|\mathcal D_{\rm bf}\rho\|_1\le4\gamma_x$, and the heating
dissipator satisfies the bound derived in Appendix~\ref{app:duhamel}. These pieces assemble into the following
statement.

\begin{theorem}[Simulation-free wall certificate]
\label{thm:wallfield}
Under all noise channels, with the thermal initial matter of the experiment,
\begin{equation}
\avg{\Phisec}(t)\le P^{\rm ref}_{\Phi}(t)+w_{\rm rest}+2\gamma_x t
+\tfrac12\int_0^t\sum_m\ndot_m\Big[\avg{\hat n_m}+\sqrt{\avg{\hat n_m^2}}+\avg{\hat n_m+1}
+\sqrt{\avg{(\hat n_m+1)^2}}\Big]_{\rm ref}\dd s ,
\label{eq:wallcert}
\end{equation}
where every quantity on the right is evaluated on the small-sector reference.
\end{theorem}

\begin{figure}
\centering
\includegraphics[width=0.98\textwidth]{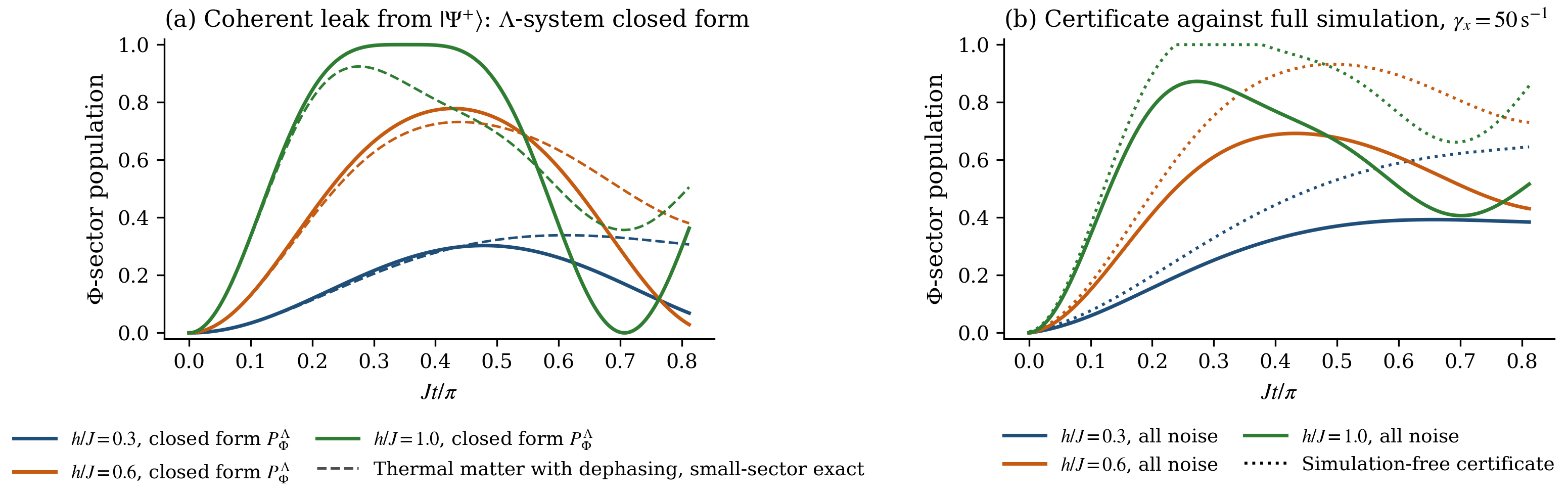}
\caption{The wall at finite field. (a) The coherent leak from $\Psip$ in the closed form of
Eq.~\eqref{eq:lambda} for three field strengths, with the small-sector exact reference for thermal matter with
motional dephasing dashed. At $h=J$ the frozen sector empties completely at $\Omega t=\pi$. (b) The
simulation-free certificate of Eq.~\eqref{eq:wallcert} against the full simulation with all noise channels and
$\gamma_x=50$ per second. The bound holds at every time point and is informative up to $h/J=0.6$.}
\label{fig:lambda}
\end{figure}

Figure~\ref{fig:lambda} shows the closed form, the small-sector reference and the certificate against the full
simulation with all noise channels and $\gamma_x=50$ per second. The certificate holds at every time point for
$h/J=0.3$, $0.6$ and $1.0$. It is informative at the first two values, with bound peaks of $0.65$ and $0.93$
against actual leaks of $0.39$ and $0.69$, and it reaches one for part of the window at $h=J$, where the leak
itself reaches $0.87$. At the end of the window the bit-flip term contributes $0.23$ and the heating term
between $0.10$ and $0.12$. The heating term is loose by about a factor of two because the Cauchy--Schwarz step
in Appendix~\ref{app:duhamel} replaces $\avg{\hat n}$ by $\sqrt{\avg{\hat n^2}}$, and the bit-flip term is the
generic one. Both could be tightened by solving the qubit-only dynamics on the same small sectors, since bit
flips also preserve the phonon number, but we have not pursued this here.

A bound that needs no reference at all also exists, but it is weaker. The commutator norm
$\|[h(\hat\sigma^{x}_{\ell_1}+\hat\sigma^{x}_{\ell_2}),\Phisec]\|=2h$ and the bit-flip relaxation $\mathcal
L^{\dagger}[\Phisec]=2\gamma_x(1-2\Phisec)$ give, by a Gr\"onwall argument,
$\avg{\Phisec}(t)\le(\tfrac12+h/2\gamma_x)(1-\ee^{-4\gamma_x t})$. This is exact at $h=0$ and reduces to $2ht$
as $\gamma_x\to0$, but it reaches one as soon as $h$ exceeds $\gamma_x$. On the chain we also use a
reference-relative form, $\avg{\Phisec}_{\rm all\ noise}\le\avg{\Phisec}_{\rm motional\ only}+2\gamma_x t$,
which follows from the same Duhamel argument with a simulated reference and which holds at both field values
tested, as Figure~\ref{fig:chain}(b) shows.

\begin{figure}
\centering
\includegraphics[width=0.98\textwidth]{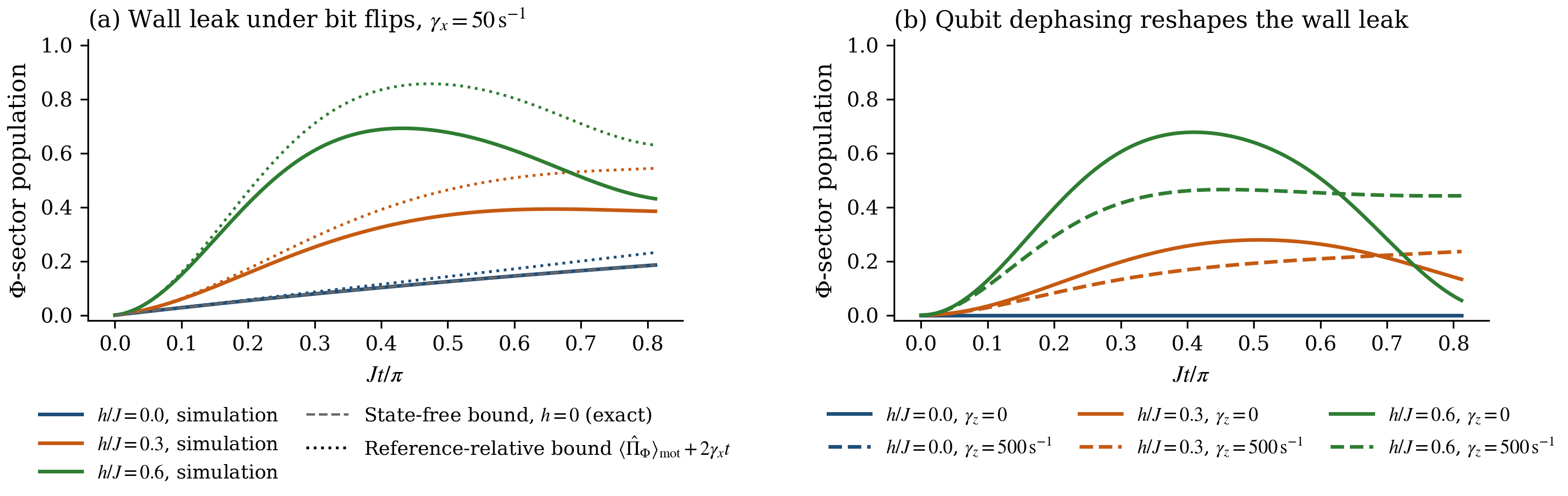}
\caption{The wall at finite field under qubit noise. (a) The leak under bit flips at $\gamma_x=50$ per second
with the reference-free bound at $h=0$ and the reference-relative bound for all three fields. (b) The coherent
leak with and without qubit dephasing at $500$ per second. Dephasing lowers the peak through the dark state
$|\Psi^{-}_{\ell_1\ell_2}\rangle$ but destroys the revival, so the late-time population settles on a plateau.}
\label{fig:dephasing}
\end{figure}

The state $|\Psi^{-}_{\ell_1\ell_2}\rangle$ is annihilated by the hopping and by the electric term alike, so
it is dark for the whole Hamiltonian. Qubit dephasing maps $\Psi^{+}$ to $\Psi^{-}$, and one might expect it
to protect the wall at finite field by parking population in the dark state. The numerical result is more
subtle. Figure~\ref{fig:dephasing} shows the $\Phi$-sector population with and without qubit dephasing at
$500$ per second for three field strengths. Dephasing lowers the peak of the coherent leak by between $15$ and
$30$ per cent, from $0.279$ to $0.236$ at $h/J=0.3$, from $0.677$ to $0.465$ at $h/J=0.6$ and from $0.909$ to
$0.660$ at $h=J$. That is the dark-state mechanism at work. Dephasing also destroys the coherent revival that
would otherwise return the population to the $\Psi$ sector, so that the late-time population settles on a
plateau instead of returning. Dephasing therefore reshapes the leak rather than suppressing it, and the
protection it offers holds for the peak and not for the time average. Motional noise alone also raises the
leak at finite field, from $0.279$ to $0.338$ at $h/J=0.3$, because the $\Lambda$-system populations depend on
the coherence between the two matter configurations, and motional dephasing destroys that coherence.

\section{Interference contrast and the separation of the channels}
\label{sec:contrast}

The third class of observable is the one in which the experiment reads off interference, namely the matter
populations and their oscillation in time. At zero electric field the Hamiltonian of the loop in the zero-flux
sector is a beam splitter between two modes, with an angle set by the gauge state, and the gauge state is
conserved. The dynamics is then linear in the bosons. For a linear open system driven by additive noise the
mean occupation at time $t$ is the noiseless mean plus a noise contribution that does not depend on the
initial state. Heating is an additive noise of this kind. It follows that at $h=0$ heating shifts
$\nbar{2}(t)$ by a smooth ramp and leaves the peak-to-trough contrast of its oscillation unchanged, for any
initial matter state. Motional dephasing is not an additive Gaussian noise. It damps the contrast at the rate
of a two-site system with dephasing on both sites, which is half the sum of the inverse coherence times.

\begin{figure}
\centering
\includegraphics[width=0.98\textwidth]{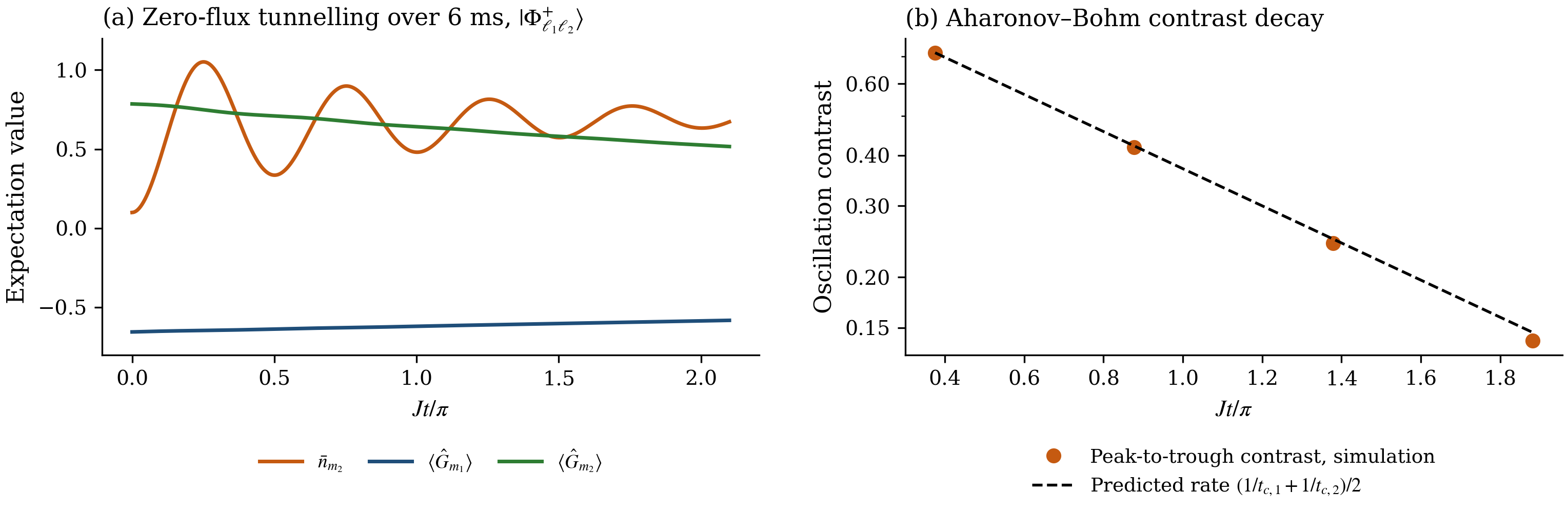}
\caption{Interference contrast at zero field on the loop over six milliseconds. (a) The population of the
second site under full motional noise, together with the two Gauss generators. (b) The peak-to-trough contrast
on a logarithmic scale against the damped two-site prediction $(1/t_{c,1}+1/t_{c,2})/2$. Heating contributes
$1$ per second to the contrast decay while destroying Gauss's law on the same run.}
\label{fig:contrast}
\end{figure}

\begin{figure}
\centering
\includegraphics[width=0.98\textwidth]{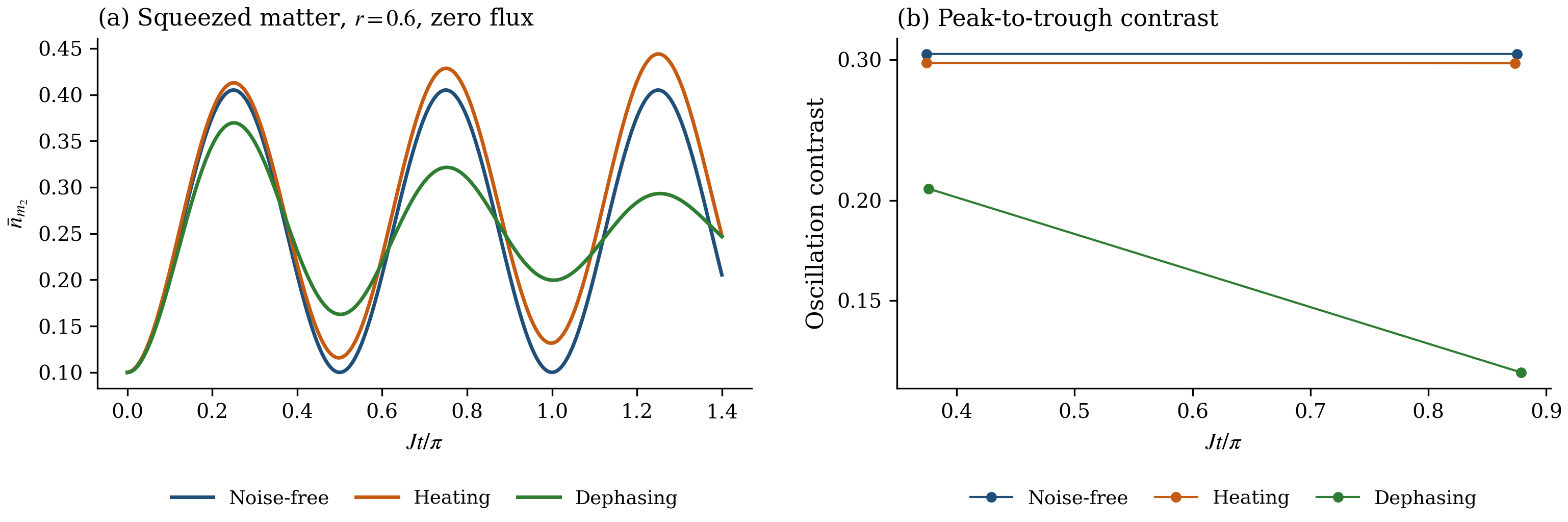}
\caption{Squeezed matter with $r=0.6$ in the zero-flux sector at $h=0$. (a) The population of the second site
without noise, under heating alone and under dephasing alone. (b) The peak-to-trough contrast. Heating leaves
the contrast at $0.999$ of its initial value and dephasing takes it to $0.59$.}
\label{fig:squeezed}
\end{figure}

Figure~\ref{fig:contrast} tests both statements over six milliseconds on the loop with the Bell-state
infidelity included. The contrast decays at $380$ per second under full motional noise, at $369$ per second
under dephasing alone and at $1$ per second under heating alone, against the prediction
$(1/t_{c,1}+1/t_{c,2})/2=369$ per second. Heating destroys Gauss's law on the same run while leaving the
contrast intact. The prediction extends to squeezed matter. With the first mode prepared in a squeezed vacuum
with $r=0.6$, which holds $0.40$ quanta, heating leaves the contrast at $0.999$ of its initial value over four
milliseconds while dephasing takes it to $0.59$, as Figure~\ref{fig:squeezed} shows. The separation of the
channels is therefore a property of linearity and not of the single-phonon sector.

Taken together with Sections~\ref{sec:gauss} and~\ref{sec:flux}, this completes the picture at zero field.
Heating breaks Gauss's law and nothing else. Motional dephasing damps the interference contrast and nothing
else. Qubit bit flips move the flux and open the wall and nothing else. Each channel owns one observable class,
and each has an exact or a first-order law in closed form. An experimenter who sees one of these observables
degrade therefore knows which channel to look at.

\begin{figure}
\centering
\includegraphics[width=0.98\textwidth]{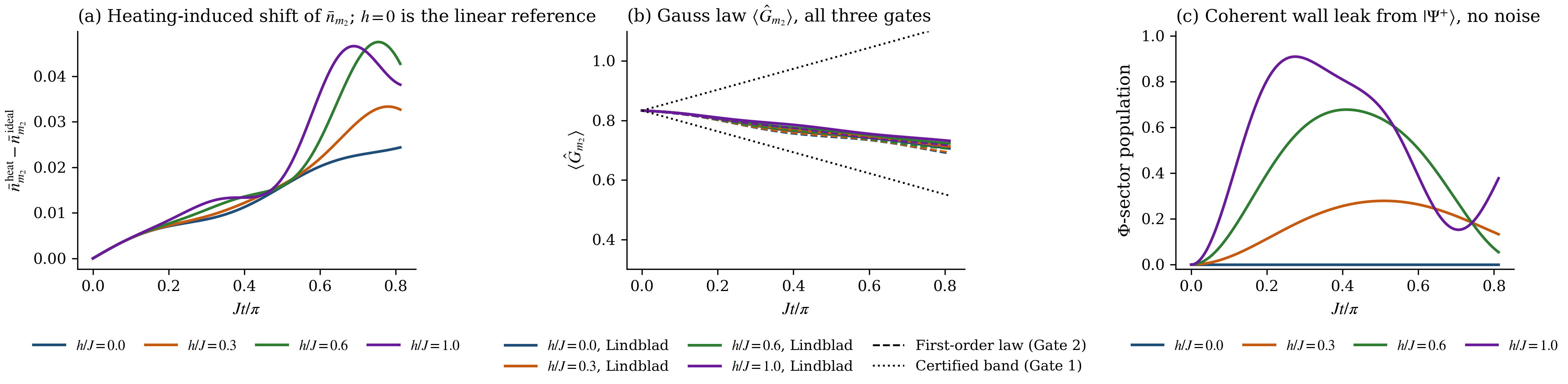}
\caption{The loop with the electric field switched on. (a) The heating-induced shift of $\nbar{2}$ for four
field strengths. At $h=0$ the shift is the smooth ramp of a linear system. With the field on it acquires
oscillatory structure and roughly doubles at $h=J$. (b) The Gauss generator on the radial mode with the
first-order law and the certified band, which do not depend on $h$. (c) The coherent leak from $\Psip$
without noise, which reaches one at $h=J$.}
\label{fig:fieldon}
\end{figure}

The separation fails when the electric field is switched on, and it fails in the heating channel. The electric
term does not commute with the hopping term, so the beam-splitter angle becomes an operator that evolves, and
the dynamics is no longer linear in the bosons. Figure~\ref{fig:fieldon} shows the heating-induced shift
$\nbar{2}^{\rm heat}-\nbar{2}^{\rm ideal}$ on the loop for four field strengths. At $h=0$ it is the smooth ramp
of a linear system. With the field on it acquires oscillatory structure locked to the $\Lambda$-system
frequencies, and by $Jt/\pi\approx0.8$ it is about twice the linear value at $h=J$. Measured as the maximum
deviation from the $h=0$ shift divided by the maximum of the $h=0$ shift, the non-additivity is $0.40$, $0.99$
and $0.99$ at $h/J=0.3$, $0.6$ and $1.0$. The effect is small in absolute terms only because the radial heating
rates are small, which is why the experiment could set it aside.

The mechanism is visible in the structure of the problem. A heating jump sends the state either into the
zero-phonon sector, where only the qubits evolve, or into the two-phonon sector, which is a two-boson $\Lambda$
scheme of its own~\cite{bazavan} with a bosonically enhanced tunnelling rate. The populations of these branches
do not add to the single-phonon dynamics in the way the linear theory requires. Two-phonon states do not exist
for hard-core matter, so the effect has no counterpart in the qubit-matter or hard-core-boson literature. A
first-order expansion in $\ndot$ around the noiseless $\Lambda$-system propagator would give the shift in
closed form. We have not carried it out, and we present the non-additivity as a quantified observation whose
analytic law remains open. The certified band of Appendix~\ref{app:duhamel} applies to it, because the
population operator is bounded on the small sectors.

\section{A calibrated acceptance gate}
\label{sec:gate}

A first-order law is a prediction, and a prediction should come with a rule that says when it may be used.
Following~\cite{ugail2026weyl}, we calibrate such a rule on an ensemble of random instances and report its
coverage on held-out instances with a finite-sample confidence interval~\cite{clopper}, in the spirit of
distribution-free risk control~\cite{bates}. The quantity gated is the first-order Gauss law of
Eq.~\eqref{eq:gausslaw} on the two-link loop. We draw $160$ loops with heating rates log-uniform between $2$
and $400$ quanta per second on each mode, coherence times uniform between $1$ and $4$ milliseconds, an electric
field uniform in $h/J\in[0,1.5]$ and a random flux sector, and we evaluate the law and the Lindblad evolution
at $t=2$ milliseconds. The declared input of the gate is the predicted fractional decay $d=1-\avg{\hat G}_{\rm
law}(t)/\avg{\hat G}(0)$, which requires no simulation, and the error is $e=|\avg{\hat G}_{\rm
sim}(t)-\avg{\hat G}_{\rm law}(t)|$. On a calibration half of the ensemble we choose the largest threshold
$d^{\star}$ such that the $97.5$ per cent quantile of $e$ among draws with $d\le d^{\star}$ lies below a
tolerance of $0.02$. The margin over the nominal $95$ per cent target guards against the finite calibration
sample. We then report the coverage on the other half.

\begin{figure}
\centering
\includegraphics[width=0.6\textwidth]{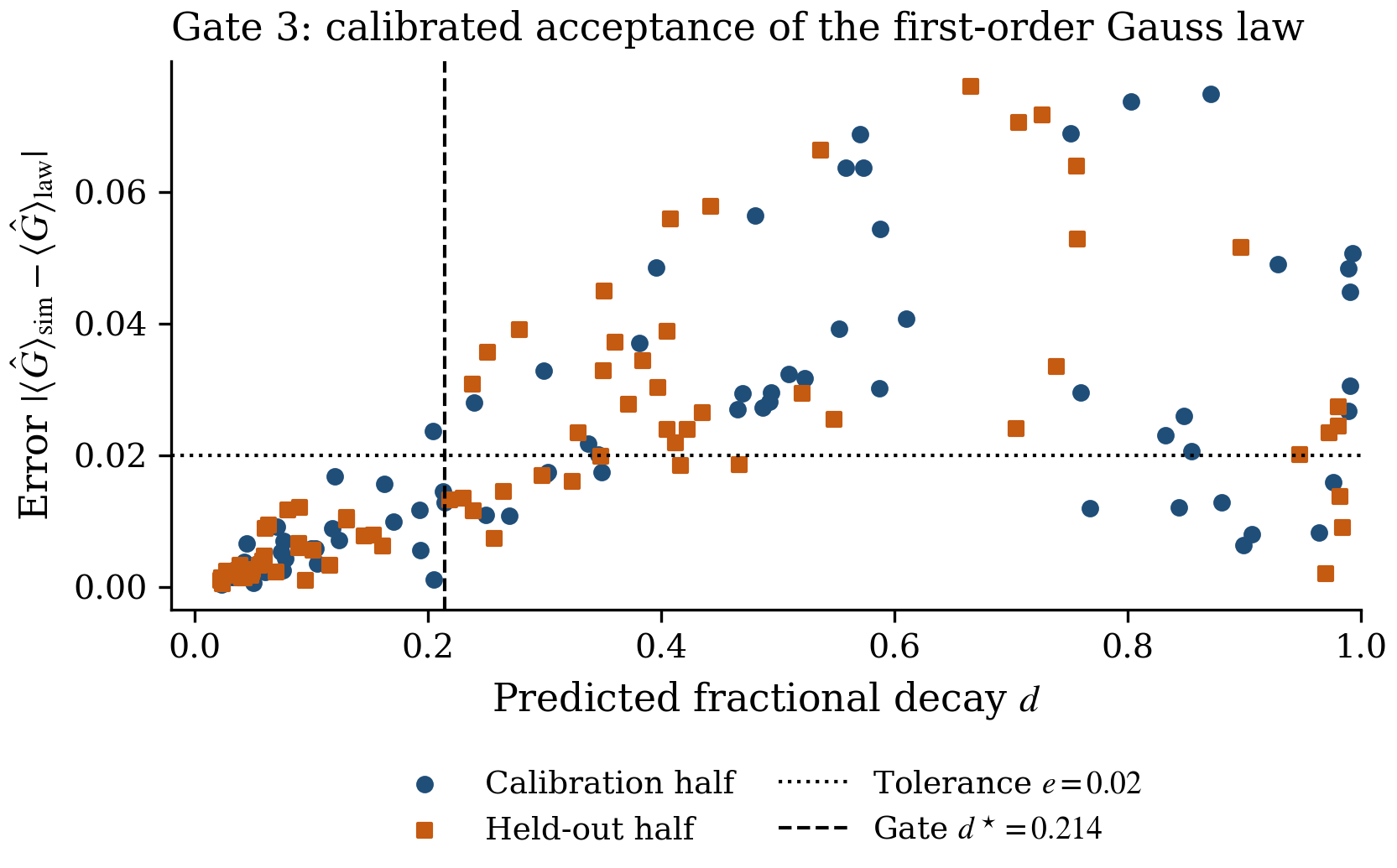}
\caption{The calibrated acceptance gate for the first-order Gauss law on $160$ random loops. The error of the
law at $t=2$ milliseconds is plotted against the predicted fractional decay, which is the declared input of
the gate. The dashed line is the calibrated threshold $d^{\star}=0.214$ and the dotted line is the tolerance
$0.02$. All $35$ accepted held-out draws lie within tolerance.}
\label{fig:gate}
\end{figure}

The calibrated gate accepts the law when $d\le0.214$. On the held-out half it accepts $35$ draws, all of which
lie within tolerance, which gives a Clopper--Pearson $95$ per cent interval of $[0.90,1.00]$ for the coverage.
The price is that $57$ per cent of the ensemble is rejected. Without the calibration margin, with the threshold
set at the $95$ per cent quantile, the held-out coverage was $0.886$, so the margin does necessary work.
Figure~\ref{fig:gate} shows the error against the declared input with the gate and the tolerance marked. The
gate is redundant where Theorem~\ref{thm:gauss} and Corollary~\ref{cor:gausscert} already apply, because the
certified bound needs no calibration. Its role here is to fix the procedure by which a law without a theorem
behind it, such as a future closed form for the heating non-additivity of Section~\ref{sec:contrast}, would be
admitted into use.

\section{Discussion}
\label{sec:discussion}

Several statements of practical use follow from the results above. The noise model stated in~\cite{saner}
cannot decay the flux correlator, so the decay reported in its supplementary figure is a measurement of a
qubit error channel that the model does not list, and the slope of that decay gives the bit-flip rate of the
device during the gauge-invariant evolution through Eq.~\eqref{eq:fluxlaw}. That rate is the most important
noise parameter for the Aharonov--Bohm protocol, because it alone controls how long a vison wall holds at zero
field.

The Gauss generators are safe on the radial modes and unsafe on the axial mode at the current heating rates,
and Corollary~\ref{cor:gausscert} gives the certified window for any mode once its heating rate and initial
occupation are known. Because the bound is superlinear in time, extending the evolution window costs more than
proportionally in gauge-symmetry breaking. The roadmap geometries of~\cite{saner} involve longer evolutions on
more modes, and they will feel this cost.

For the interference itself, the contrast at zero field is limited by motional dephasing alone and is immune
to heating for any matter state, so squeezed matter may be used freely at zero field. Once the electric field
is on, the heating shift of the populations is no longer the additive ramp that a linear model would subtract,
and the frozen sector leaks coherently at the rate set by Eq.~\eqref{eq:lambda}, with the complete and periodic
emptying at $h=J$ as a sharp test. The certificate of Theorem~\ref{thm:wallfield} gives the guaranteed envelope
within which the leak must lie for given noise rates.

The study of gauge violation under errors in quantum simulators was placed on a quantitative footing by
Halimeh and Hauke~\cite{halimeh2020reliability}. They showed that coherent gauge-breaking terms of strength
$\lambda$ produce a violation that grows as $\lambda^2t^2$ from a gauge-invariant state and then enters a
staircase of prethermal plateaus. Halimeh, Kasper and Hauke~\cite{halimeh2020fate} then showed that Markovian
decoherence produces a violation that grows as $\gamma t$, and they identified the violation with a
mean-square displacement across gauge sectors. The companion paper~\cite{halimeh2020crossover} derived these
scalings in time-dependent perturbation theory for general local and global symmetries, treated particle loss
on hard-core bosons explicitly, and showed that matter dephasing contributes no violation at short times. Our
Theorem~\ref{thm:gauss} is the exact form of their incoherent law for soft bosonic matter under heating, valid
to all orders, and its occupation factor is the feature that their hard-core model excludes. Gauge protection
by single-body terms~\cite{halimeh2021protection} and by engineered dissipation~\cite{stannigel} addresses
coherent errors and is complementary to the present analysis, which concerns the incoherent channels of a
specific device.

The hybrid qubit--oscillator encoding was proposed and analysed by B\u{a}z\u{a}van and
co-workers~\cite{bazavan}, whose exact one-boson Rabi solution, two-boson $\Lambda$ scheme and plaquette
analysis we use throughout, and it was realised by Saner and co-workers~\cite{saner}, following earlier
proposals for synthetic gauge fields on phonons~\cite{bermudez2011} and for hybrid analog--digital simulation
with controlled phonon--ion interactions~\cite{davoudi}. The confinement of charges by $\pi$-flux loops in a
$\ZZ$ loop gauge theory, and the dynamical Aharonov--Bohm cages that result, were analysed by Domanti,
Bermudez and Amico~\cite{domanti} in the coherent and finite-density setting. Our Theorems~\ref{thm:wall}
and~\ref{thm:wallfield} supply the noise robustness of those cages. The three-tier certification architecture
and the quantile-calibrated gate come from our work on Weyl diagnostics in numerical
relativity~\cite{ugail2026weyl}, and the idea of predicting decay from the noise channel description comes from
our work on noisy quantum neural networks~\cite{ugail2026access,ugail2026coherence}.

The results rest on a Lindblad model with the parameters the experiment reports, and the most consequential
modelling assumption is the form of motional dephasing. We have modelled it as a Markovian process with an
exponential loss of coherence at the quoted coherence times. If the motional dephasing of the device is
dominated by slow drifts of the trap frequency, the loss of coherence is closer to Gaussian in time, and the
contrast decay of Section~\ref{sec:contrast} changes shape, though not its attribution to the dephasing
channel. The heating model is the standard infinite-temperature bath, which is appropriate for electric-field
noise on a trap electrode~\cite{turchette,brownnutt}.

The theorems are exact statements about this model and do not depend on the geometry. The numerical tests have
been carried out at modest Fock truncations on a single processor core, with truncation four on the chain and
twelve on the squeezed-matter runs, and they should be repeated at higher truncations before any number is used
in a comparison with measured data. The tetrahedron of the roadmap of~\cite{saner}, with four sites and six
links, has not been attempted, because its Hilbert space calls for quantum trajectories rather than density
matrices. The calibration ensemble of Section~\ref{sec:gate} is small, and its coverage interval is
correspondingly wide.

Two analytic gaps remain. The heating term of the wall certificate is loose by about a factor of two and the
bit-flip term is the generic one, and both could be tightened by solving the qubit-only dynamics on the small
phonon-number sectors. More substantially, the non-additive heating shift of the populations at finite field
has been quantified and bounded but not derived. Its derivation is a first-order expansion in $\ndot$ around
the noiseless $\Lambda$-system propagator in which each jump branch is itself exactly solvable, and it is the
natural next step, because it would be the first analytic statement in this programme that has no counterpart
for hard-core matter.

\section{Conclusions}
\label{sec:conclusions}

In this work, we have asked whether the noise of a hybrid qubit--oscillator quantum simulator of a $\ZZ$
lattice gage theory can be predicted rather than only simulated, and we have answered in the affirmative for
each of the three observable classes that the experiment measures. The Gauss generators obey an exact equation
of motion under motional heating whose rate grows with the occupation of the mode, so that the violation of
Gauss's law accelerates in time. That equation yields a certified bound through the exactly known total phonon
number and a first-order law from the noiseless populations, and both hold on every geometry and at every
electric field we tested. The flux obeys an exact exponential law under qubit bit flips and is untouched by
every other channel.

At zero electric field the three channels of the device act on three different observables. Heating breaks
Gauss's law and leaves the flux and the interference contrast intact for any matter state, including squeezed
matter. Motional dephasing damps the contrast at the damped two-site rate and leaves Gauss's law and the flux
intact. Qubit bit flips alone move the flux and alone open a $\pi$-flux loop, with an exact law for each. One
consequence is that the decay of the flux correlator reported by the experiment cannot be produced by the noise
model it lists, and is therefore a measurement of its bit-flip rate.

With the electric field on, the frozen sector leaks coherently with a closed form on a three-level system,
including a complete and periodic emptying when the field equals the tunnelling, and a simulation-free
certificate for the leak follows from the exact decomposition of the dephasing dynamics into small
phonon-number sectors. The separation of the channels fails in the heating sector. There the heating shift of
the populations becomes non-additive at the level of the heating effect itself, through two-phonon branches
that hard-core matter does not possess. We have quantified and bounded this effect and left its closed form
open.

The architecture of certified bound, predictive law and calibrated gate has carried over from spectral
diagnostics in numerical relativity to lattice gauge theory without modification, and we expect it to carry to
other platforms in which the noise channels are known and the observables are gauge-invariant. The natural
next steps are the derivation of the non-additive heating law, the tightening of the wall certificate on the
small sectors, the extension to the tetrahedron by quantum trajectories, and the comparison with the measured
curves of the experiment, which would fix the one noise parameter the experiment did not report.

\section*{Acknowledgments}
The author would like to thank the Centre for Visual Computing and Intelligent Systems at the University of
Bradford for providing the computing resources used in this work.

\section*{Funding}
No specific funding was received for this work.

\appendix

\section{Duhamel bounds}
\label{app:duhamel}

Let the Lindbladian be $\mathcal L=\mathcal L_0+\mathcal D$ with $\mathcal D\rho=\sum_k(\hat L_k\rho\hat
L_k^{\dagger}-\tfrac12\{\hat L_k^{\dagger}\hat L_k,\rho\})$, and let $\rho_t$ and $\rho^{(0)}_t$ be the states
evolved under $\mathcal L$ and under $\mathcal L_0$ from the same initial state. The Duhamel identity reads
$\rho_t-\rho^{(0)}_t=\int_0^t\ee^{(t-s)\mathcal L}\mathcal D\rho^{(0)}_s\dd s$. Since $\ee^{\tau\mathcal L}$ is
completely positive and trace preserving, it does not increase the trace norm of a Hermitian
operator~\cite{nielsen}, so that $\|\rho_t-\rho^{(0)}_t\|_1\le\int_0^t\|\mathcal D\rho^{(0)}_s\|_1\dd s$. For a
single jump operator, $\|\hat L\rho\hat L^{\dagger}\|_1=\mathrm{tr}(\hat L^{\dagger}\hat L\rho)=\avg{\hat
L^{\dagger}\hat L}$ because the operator is positive, and for $\hat A=\hat L^{\dagger}\hat L$ the
Cauchy--Schwarz inequality gives $\|\hat A\rho\|_1=\|\hat A\rho^{1/2}\rho^{1/2}\|_1\le\|\hat
A\rho^{1/2}\|_2\|\rho^{1/2}\|_2=\sqrt{\avg{\hat A^2}}$. Hence
\begin{equation}
\|\mathcal D\rho\|_1\le\sum_k\Big[\avg{\hat L_k^{\dagger}\hat L_k}+\sqrt{\avg{(\hat L_k^{\dagger}\hat L_k)^2}}\Big].
\label{eq:duhamel}
\end{equation}
For the heating jumps $\sqrt{\ndot}\,\hat a$ and $\sqrt{\ndot}\,\hat a^{\dagger}$ the operators $\hat
L^{\dagger}\hat L$ are $\ndot\hat n$ and $\ndot(\hat n+1)$, which gives the integrand of
Eq.~\eqref{eq:wallcert}. For a bit flip $\sqrt{\gamma_x}\hat\sigma^{x}$ one has
$\|\hat\sigma^{x}\rho\hat\sigma^{x}-\rho\|_1\le2$ directly, so two links contribute $4\gamma_x$. For any
observable $\hat O$ with extreme eigenvalues $\lambda_{\min}$ and $\lambda_{\max}$,
$|\mathrm{tr}\,\hat O(\rho_t-\rho^{(0)}_t)|\le\tfrac12(\lambda_{\max}-\lambda_{\min})\|\rho_t-\rho^{(0)}_t\|_1$.
This is half the trace distance for a projector, and it is $\tfrac12N_{\max}$ times the trace distance for a
population operator restricted to sectors with at most $N_{\max}$ phonons.

\section{Small-sector decomposition and the $\Lambda$ system}
\label{app:sectors}

The hopping term of Eq.~\eqref{eq:H} conserves the total phonon number, the electric term acts on the qubits
alone, and motional dephasing is a function of the phonon numbers. The Lindbladian without heating therefore
commutes with $\hat N$ and maps each sector of fixed $N$ into itself. A matter state that is diagonal in the
Fock basis, such as the thermal and sideband-prepared states of the experiment, is a mixture of Fock states
each of which lies in one sector. Because the evolution is linear, the mixture evolves as the weighted sum of
the sector evolutions with no cross terms. The sector with $N$ phonons on two sites and two links has dimension
$4(N+1)$ before the gauge sector is fixed, and the dynamics from a $\Psi^{+}$ gauge state explores a smaller
reachable subset. For $N=1$ the reachable subset is the three-state $\Lambda$ system of
Section~\ref{sec:wallfield}, with Hamiltonian
\begin{equation}
\hat H_{\Lambda}=\begin{pmatrix}0&2h&0\\2h&0&2J\\0&2J&0\end{pmatrix},
\end{equation}
whose eigenvectors are $(2J,0,-2h)/\Omega$ with eigenvalue $0$ and $(2h,\pm\Omega,2J)/(\sqrt2\,\Omega)$ with
eigenvalues $\pm\Omega$. Expanding the initial state $(1,0,0)$ in this basis gives the amplitudes
$b(t)=-\ii(2h/\Omega)\sin\Omega t$ and $c(t)=(4hJ/\Omega^2)(\cos\Omega t-1)$ on the two $\Phi$-sector states,
and their squared moduli sum to Eq.~\eqref{eq:lambda}. For $N=0$ the dynamics is a rotation of the gauge qubits
alone, and for $N=2$ the reachable subset has five states. The reference of Theorem~\ref{thm:wallfield}
evolves each sector with $N\le3$ exactly, with the motional dephasing included, and the phonon moments needed
for the heating term of Eq.~\eqref{eq:wallcert} are read off the same evolution.

\end{document}